\documentclass[twocolumn]{aastex6}

\usepackage{amsmath}
\usepackage{units}

\renewcommand*{\vec}[1]{\boldsymbol{#1}}
\newcommand{\grad}{\vec{\nabla}}
\renewcommand{\dot}{\vec{\cdot}}
\renewcommand*{\emph}[1]{\textit{\textbf{#1}}}

\newcommand{\MESA}{\textsc{MESA}}
\newcommand{\Msun}{\ensuremath{\mathrm{M}_{\sun}}} 

\newcommand{\Dturb}{\ensuremath{D_{\rm t}}} 

\begin{document}
\label{firstpage}
\shorttitle{Convectively Bounded Flames}
\title{Turbulent Chemical Diffusion in Convectively Bounded Carbon Flames}

\shortauthors{Lecoanet et al.}
\author{Daniel Lecoanet\altaffilmark{1,2}, Josiah Schwab\altaffilmark{1,2}, Eliot Quataert\altaffilmark{1,2}, Lars Bildsten\altaffilmark{3,4}, F. X. Timmes\altaffilmark{3,5}, Keaton J. Burns\altaffilmark{6},
Geoffrey M. Vasil\altaffilmark{7}, Jeffrey S. Oishi\altaffilmark{8}, \& Benjamin P. Brown\altaffilmark{9}}
\altaffiltext{1}{Physics Department, University of California, Berkeley, CA 94720, USA}
\altaffiltext{2}{Astronomy Department and Theoretical Astrophysics Center, University of California, Berkeley, CA 94720, USA}
\altaffiltext{3}{Kavli Institute for Theoretical Physics, University of California, Santa Barbara, CA 93106, USA}
\altaffiltext{4}{Department of Physics, University of California, Santa Barbara, CA 93106, USA}
\altaffiltext{5}{School of Earth and Space Exploration, Arizona State University, Tempe, AZ 85287, USA}
\altaffiltext{6}{Department of Physics, Massachusetts Institute of Technology, Cambridge, Massachusetts 02139, USA}
\altaffiltext{7}{School of Mathematics \& Statistics, University of Sydney, NSW 2006, Australia}
\altaffiltext{8}{Department of Physics \& Astronomy, Bates College, Lewiston, ME 04240}
\altaffiltext{9}{Laboratory for Atmospheric and Space Physics and Department of Astrophysical \& Planetary Sciences, University of Colorado, Boulder, Colorado 80309, USA}

\email{dlecoanet@berkeley.edu}


\begin{abstract}
It has been proposed that mixing induced by convective overshoot can disrupt the inward propagation of carbon deflagrations in super-asymptotic giant branch stars.  To test this theory, we study an idealized model of convectively bounded carbon flames with 3D hydrodynamic simulations of the Boussinesq equations using the pseudospectral code Dedalus.  Because the flame propagation timescale is much longer than the convection timescale, we approximate the flame as fixed in space, and only consider its effects on the buoyancy of the fluid.  By evolving a passive scalar field, we derive a {\it turbulent} chemical diffusivity produced by the convection as a function of height, $\Dturb(z)$.  Convection can stall a flame if the chemical mixing timescale, set by the turbulent chemical diffusivity, $\Dturb$, is shorter than the flame propagation timescale, set by the thermal diffusivity, $\kappa$, i.e., when $\Dturb>\kappa$.  However, we find $\Dturb<\kappa$ for most of the flame because convective plumes are not dense enough to penetrate into the flame.  Extrapolating to realistic stellar conditions, this implies that convective mixing cannot stall a carbon flame and that ``hybrid carbon-oxygen-neon'' white dwarfs are not a typical product of stellar evolution.
\end{abstract}


\section{Introduction}
\label{intro}
Super-asymptotic giant branch (SAGB) stars are characterized by the
development of a degenerate carbon-oxygen (CO) core and the subsequent
ignition of off-center carbon fusion within it.  Stellar evolution
calculations show that this occurs in stars that have zero-age main
sequence masses $\approx 7-11\,\Msun$, with this mass range depending
on the metallicity and on modeling assumptions such as the mass loss
rate and the efficiency of mixing at convective boundaries.  Carbon
ignition initially occurs as an off-center flash, but after one or
more of these flashes, a self-sustaining carbon-burning front can
develop \citep[see e.g.,][]{Siess06, Farmer15}.  This ``flame''
propagates towards the center of the star extremely sub-sonically, as
heat from the burning front is conducted inward.  The heat from the
burning also drives a convective zone above the burning front, and in
the quasi-steady-state, the energy released by carbon fusion is
balanced by energy losses via neutrino cooling in this convective zone
\citep{Timmes94}.  As the carbon-burning flame propagates to the
center, it leaves behind oxygen-neon (ONe) ashes.  This process
creates the core that will become a massive ONe WD or collapse to a
neutron star, powering an electron-capture supernova \citep{Miyaji80}.

However, the presence of additional mixing near the flame can lead to
its disruption, preventing carbon burning from reaching the center.
There are at least two physical processes that may play a role in this
region: (1) mixing driven by the thermohaline-unstable configuration
of the hot ONe ash on top of the cooler CO fuel and (2) mixing driven
by the presence of a convective zone above the flame via convective
overshoot.  These processes were investigated by
\citet{Denissenkov13b} using 1D stellar evolution models.  With a
thermohaline diffusion coefficient informed by multi-dimensional
hydrodynamics simulations, they concluded that thermohaline mixing was
not sufficient to disrupt the flame.  However, they did find that the
introduction of sufficient convective boundary mixing---using a model
of exponential overshooting \citep{Freytag96, Herwig00}---disrupted the flame, preventing carbon burning from reaching the
center.  This led to the production of ``hybrid C/O/Ne'' WDs, in which
a CO core is overlaid by an ONe mantle.  Several groups have begun to
model the explosions that would originate from objects with this
configuration \citep{Denissenkov15, Kromer15, Bravo16, Willcox16}.

%

Is mixing sufficiently vigorous to disrupt the carbon flame?  This is
a key question for understanding the final outcomes of SAGB stars and
the WDs they produce.  If the thermal diffusivity $\kappa$ is much
larger than the chemical diffusivity $D$, the flame propagates into
fresh fuel much more quickly than the fuel and ash can mix, allowing
the flame to successfully propagate to the center of the star.  We
estimate $\kappa/D\sim 10^6$ using the thermal conductivity in MESA
(which is drawn from \citealt{Cassisi07}) and a chemical diffusivity
from \citet{Beznogov14}.  However, convective mixing could produce a
{\it turbulent} diffusivity $\Dturb$, which if similar to $\kappa$,
could mix ash into the fuel, stalling the flame, as was found in
\citet{Denissenkov13b}.

In this paper, we present 3D simulations of an idealized model of a
convectively-bounded carbon flame.  These simulations allow us to
measure the enhanced mixing due to convective overshoot, and to determine if $\Dturb>\kappa$ within the flame.  Section~\ref{sec:carb-flame-prop} summarizes the properties of carbon
flames, which we use to motivate the problem setup presented in
Section~\ref{sec:prob-setup}.  Section~\ref{sec:results} presents the
results of our simulations and we discuss their implications in
Section~\ref{sec:conclusions}.

\section{Carbon Flame Properties}
\label{sec:carb-flame-prop}

\begin{figure}
\includegraphics[width=\columnwidth]{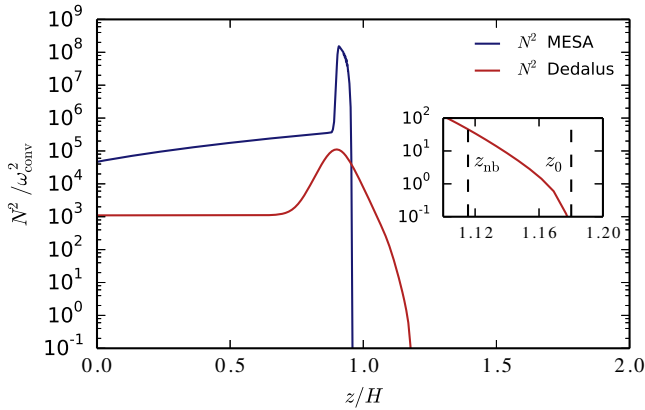}
  \caption{The blue line shows the buoyancy frequency squared near a carbon flame from a 9.5 $\Msun$ star evolved in MESA.  The red line is the buoyancy frequency squared from the Dedalus simulation R8 (very close to its initial profile, see equation~\ref{eqn:N_0^2 dedalus}).  Due to computational limitations, the buoyancy frequency in the model of the carbon flame is much lower and the transition between the buoyancy peak and the convective region is much more gradual, in Dedalus than in the MESA model. These differences both act to enhance the convective mixing via overshoot in Dedalus.  The inset shows the neutral buoyancy height $z_{\rm nb}$ and the bottom of the convection zone $z_0$ in the Dedalus simulation.  In the MESA model, this region is not resolved, with a width $z_0 - z_{\rm nb} < 3 \times 10^{-3}H$.}\label{fig:buoyancy frequency}
\end{figure}

To obtain an example of the structure of a carbon flame, we evolve a
star with zero-age main sequence mass of 9.5 $\Msun$ using revision
6794 of the \MESA\ stellar evolution code\footnote{\MESA\ is available
  at \url{http://mesa.sourceforge.net/}.} \citep{Paxton11, Paxton13,
  Paxton15}.  We used the publicly available inlists of
\cite{Farmer15}, who undertook a systematic study of carbon flames in
SAGB stars.  We did not include the effects of overshoot at the
convective boundaries, but did include the effects of thermohaline
mixing.  The Brunt-V\"{a}is\"{a}l\"{a} (buoyancy) frequency profile of
the carbon flame is shown by the blue line in Fig.~\ref{fig:buoyancy
  frequency}.  The thermal component dominates the buoyancy frequency.
The much smaller compositional component is destabilizing, but
\citet{Denissenkov13b} found thermohaline mixing to not affect flame
propagation.  The flame structure in Fig.~\ref{fig:buoyancy frequency}
is similar to that shown in Figure~3 of \citet{Denissenkov13b}.


The peak of the buoyancy frequency profile shown in
Fig.~\ref{fig:buoyancy frequency} is at a Lagrangian mass coordinate
of $M_r = 0.13\,\Msun$.  The properties of the flame change as it
propagates, but the following numbers are representative throughout
the evolution.  The inward flame velocity is
$u = \unit[9\times10^{-4}]{cm\,s^{-1}}$; it will take
$\sim \unit[10^{4}]{yr}$ to reach the center.  The flame width,
$\delta$, measured in terms of pressure scale height,
$H = \unit[2\times10^8]{cm}$, is $\delta / H \approx 0.03$.  The
timescale for the flame to cross itself,
$t_{\mathrm{cross}} = \delta / u \approx \unit[200]{yr}$, which is
also the timescale for the nuclear burning to occur.  The convection
zone above the flame has a radial extent of about one pressure scale
height and a convective turnover timescale of a few hours.
This implies that there are $\sim 10^5$ convective turnover times in
the time it takes flame to cross itself.  Thus, over the relatively
smaller number of convective turnover times covered by our
simulations, $\sim 10^2$, the flame is effectively stationary,
allowing us to exclude nuclear reactions in our model.

We note that our stationarity assumption is not universally
applicable.  Convectively bounded oxygen-neon-burning flames, which
can also occur in the late evolution of stars in this mass range are
thinner, $\delta \sim \unit[10^3]{cm}$, and have higher velocities,
$u \sim \unit[1]{cm\,s^{-1}}$, as a result of the higher energy
generation rate \citep{Timmes94, Woosley15}. Consequently, the time
for the flame to traverse its width may be $\lesssim 10$ convective
turnover times.  Thus it is difficult to anticipate how our
simulations carry over to the case of oxygen-neon flames.

The Mach number of the convection is $\approx 4 \times 10^{-5}$, so
compressibility does not play an important role in the convection.  To
measure the degree of turbulence of the convection, we calculate the
Rayleigh number
\begin{align}\label{eqn:rayleigh}
{\rm Ra} = \frac{\omega_0^2 H^4}{\nu\kappa},
\end{align}
which is the ratio of convective driving to diffusive damping.  The
variables $\omega_0$ and $H$ represent typical convective frequencies
and lengths, and $\nu$ and $\kappa$ are the kinematic viscosity and
thermal diffusivity.  We estimate the convection driven by a carbon
flame to have ${\rm Ra} \sim 10^{24}$, using
$\omega_0 \sim \unit[3\times10^{-4}]{s^{-1}}$, $H \sim \unit[2\times10^{8}]{cm}$, 
$\nu \sim \unit[5\times10^{-2}]{cm^2\,s^{-1}}$ \citep{Itoh83} and
$\kappa \sim \unit[3 \times 10^3]{cm^2\,s^{-1}}$ \citep{Itoh87}.  This large Rayleigh number means the flow is extremely turbulent.

Flames maintain coherence because their thermal diffusivity is much larger than their chemical diffusivity.  The ratio of these diffusivities is the Lewis number
\begin{align}\label{eqn:lewis}
{\rm Le} = \frac{\kappa}{D}.
\end{align}
For carbon flames, we estimate ${\rm Le} \sim 10^6$.

\section{Problem Setup}
\label{sec:prob-setup}

Our idealized simulations make a variety of assumptions to render this problem computationally tractable.  We do not include nuclear reactions because the flame is effectively stationary on the convection time scale.  We use the Boussinesq approximation because the Mach number of the convection is small, and the height of the convection zone is about a scale height, so we do not believe density contrasts across the convection zone will strongly alter the dynamics.

\subsection{Equations, Numerics, \& Assumptions}

We solve the 3D Boussinesq equations \citep{SpiegelVeronis60} using the Dedalus\footnote{Dedalus is available at \url{http://dedalus-project.org}.} pseudo-spectral code \citep{burns17}.
\begin{align}
\partial_t\vec{u} + \grad p - \nu\nabla^2\vec{u} - g T\vec{e}_z & = -\vec{u}\dot\grad\vec{u}, \\
\partial_t T - \kappa \nabla^2 T & = -\vec{u}\dot\grad T + \bar{H}, \\
\grad\dot\vec{u} & = 0,
\end{align}
where $\vec{u}$ and $p$ are the fluid velocity and pressure, respectively,  $T$ is the temperature normalized to a reference value,  $g$ is the gravitational acceleration, and $\vec{e}_z$ is the unit vector in the vertical direction. We neglect the compositional effects on buoyancy (and thus thermohaline mixing), and always use $\nu=\kappa$ for computational convenience. 

Convective overshoot is particularly sensitive to the buoyancy frequency profile \citep[e.g.,][]{brummell02}.  Thus, we study convective overshoot using a buoyancy frequency profile inspired by a carbon flame.  This assumes that the most important property affecting turbulent mixing of a carbon flame is its strong buoyancy stabilization.

The simulations are initialized with a temperature profile $T_0(z)$ satisfying $N_0^2(z)=g {\rm d} T_0/{\rm d}z$, where
\begin{align}\label{eqn:N_0^2 dedalus}
N_0^2 = -\omega_0^2 +& N^2_{\rm tail} \frac{1}{2}\left[1-\tanh\left(\frac{z-z_{\rm fl}}{\Delta z_{\rm fl}}\right)\right] \nonumber \\ +& N^2_{\rm fl} \cosh\left(\frac{z-z_{\rm fl}}{\Delta z_{\rm fl}}\right)^{-2},
\end{align}
where $\omega_0^2$ is a characteristic convective frequency, and we take $N^2_{\rm tail}=100\omega_0^2$, $N^2_{\rm fl}=10^4\omega_0^2$ as approximations to the MESA model.  The position of the buoyancy peak (``flame'') is $z_{\rm fl}=0.9 H$ and its half-width is $\Delta z_{\rm fl}=0.05 H$, where $H$ represents a pressure scale height.  We plot the time-averaged buoyancy frequency profile of simulation R8 in Fig.~\ref{fig:buoyancy frequency} with a red line.  All simulations have very similar buoyancy frequency profiles, which differ from $N_0^2$ only very close to the bottom of the convection zone.  We also include a heating term $\bar{H}=-\kappa \partial_z^2 T_0$ which exactly balances the diffusion of $T_0$.  This maintains the buoyancy profile and convection over the course of our simulations, enforcing the stationary assumption.

It is important to note that a flame with the width and thermal diffusivity used in our simulations would propagate across itself in only $10^{1-2}$ convective turnover times.  This is because the thermal diffusion in the simulations is much more rapid than in a star.  As a result, the stationary
buoyancy peak in our simulations does not self-consistently represent
a real carbon flame, whose properties would depend on the thermal
diffusivity.  However, in the limit in which the thermal diffusivity
in the simulation approaches the thermal diffusivities realized in
stars, the simulations would provide a good approximation to
convective overshoot in real carbon flames.  Therefore, we hold the
buoyancy profile of the model ``flame'' fixed as we carry out
simulations with different microphysical diffusivities.  We show below
that despite the need to extrapolate the simulation results, we can
nonetheless draw firm conclusions about convective mixing in carbon
flames.

The simulations are non-dimensionalized using the pressure scale height $H$, and the initial buoyancy frequency in the convection zone $|N_0(z=2H)|=\omega_0$.  These are used to define a Rayleigh number (Eqn.~\ref{eqn:rayleigh}).  The limited resolution of any multi-dimensional astrophysics simulation requires diffusivities much larger than in stars, so we can only reach ${\rm Ra}=10^9\ll 10^{24}$.  Our highest resolution simulation required about 3 million cpu-hours on the Pleiades supercomputer.

We define the bottom of the convection zone, where $N^2=0$, to be $z_0$.  We also define the height of neutral buoyancy $z_{\rm nb}$, the point at which $\langle T(z_{\rm nb})\rangle_{x,y,t}=\langle T(z_{\rm top})\rangle_{x,y,t}$, where $\langle \cdot \rangle_{x}$ denotes an average over $x$, and $z_{\rm top}$ is the top of the domain (see inset in Fig.~\ref{fig:buoyancy frequency}).  Plumes emitted at the top of the convection zone become neutrally buoyant at $z_{\rm nb}$.  Convective plumes cross $z_0$, but rarely pass below $z_{\rm nb}$.

The convection frequency $\omega_{\rm conv}$ and the height of the convection zone $H_{\rm conv}$ are outputs of the simulation.  We define $H_{\rm conv}$ using $z_0$ and
\begin{align}
\omega_{\rm conv} = 2\pi\frac{w_{\rm rms}}{H_{\rm conv}},
\end{align}
where $w_{\rm rms}$ is the root-mean-square vertical velocity in the convection zone.  We find $H_{\rm conv}\approx 0.83 H$ and $\omega_{\rm conv}\sim 0.3 \omega_0$.

Simulations with higher ${\rm Ra}$ have smaller $\omega_{\rm conv}$.  This is driven by the thermal equilibration of the system.  In statistically steady state, the convection zone is almost isothermal, so the temperature perturbation at the bottom of the convection zone is about $-H_{\rm conv}\omega_0^2/g$.  To satisfy our bottom boundary condition, the stable region has a temperature gradient of about $-H_{\rm conv}\omega_0^2/(g H_{\rm stable})$, where $H_{\rm stable}=2-H_{\rm conv}$.  Because the temperature gradient in the stable region is independent of $\kappa$, the heat flux scales like $\kappa\sim{\rm Ra}^{-1/2}$.  To maintain flux balance, this heat flux must be carried by the convective flux in the convection zone, which scales like $w_{\rm rms}^{3}$.  Thus, we have that $w_{\rm rms}\sim \omega_{\rm conv}\sim {\rm Ra}^{-1/6}$.

Plumes become neutrally buoyant at $z_{\rm nb}$, but will penetrate further due to their inertia.  To measure this effect, we define an ``overshoot number'' ${\rm Ov}$, which is the ratio of inertial to buoyancy forces near $z_{\rm nb}$,
\begin{align}\label{eqn:Ov}
{\rm Ov} \equiv \frac{\omega_{\rm conv}^2}{N_{\rm fl}^2}\frac{\Delta z_{\rm fl}}{H},
\end{align}
where we estimate the inertia of the fluid as $\sim \omega_{\rm conv}^2H$, and the buoyancy as $H^2 N_{\rm fl}^2/\Delta z_{\rm fl}$.  The latter assumes the derivative of the buoyancy frequency squared near $z_{\rm nb}$ is proportional to $N_{\rm fl}^2/\Delta z_{\rm fl}$.  We report ${\rm Ov}$ for our simulations in Table~\ref{tab:sims}.

For comparison, we estimate real flames have ${\rm Ov}\sim 10^{-10}$, using $N_{\rm fl}^2\sim 2\times 10^8$ and $\Delta z_{\rm fl}=0.03H$.  However, the buoyancy frequency profile is actually much steeper than this linear estimate, so the real ${\rm Ov}$ is likely even smaller (see Fig.~\ref{fig:buoyancy frequency}).  Our chosen buoyancy profile differs from the MESA model in two important ways: (1)~the peak is at lower frequencies; and (2)~the buoyancy frequency approaches zero more gradually.  This is necessary because it is difficult to resolve the fast buoyancy timescale, and sharp buoyancy gradients numerically.  Both these changes lead to substantially higher ${\rm Ov}$ than we expect in real flames.  Thus, we expect our simulated plumes to penetrate much further than the convective plumes driven by carbon flames.  Table~\ref{tab:sims} also reports the Reynolds number, a measure of the degree of turbulence in the fluid, defined as
\begin{align}\label{eqn:reynolds}
{\rm Re} = \frac{w_{\rm rms}H_{\rm conv}}{\nu}.
\end{align}

We solve the equations in cartesian geometry ($x,y,z$), in the domain $[0,4H]^2\times [0,2H]$. The simulations are periodic in the horizontal directions, and no-slip with zero temperature perturbation at the top and bottom.  All quantities are expanded in a Fourier series in the horizontal directions.  In the vertical direction, quantities are independently expanded in Chebyshev polynomials over the domain $[0, 1.05H]$, and over the domain $[1.05H, 2H]$, with boundary conditions imposed at $z=1.05H$ to maintain continuity of each quantity and its first vertical derivative.  An equal number of Chebyshev modes are used in each vertical sub-domain.  3/2 dealiasing is used in each direction.  We use mixed implicit-explicit timestepping, where all the linear terms are treated implicitly, and the remaining terms treated explicitly.  The timestep size is determined using the Courant--Friedrichs--Lewy (CFL) condition.  Table~\ref{tab:sims} describes the simulations presented in this paper.

\begin{table*}
  \caption{List of simulations.  The Rayleigh and Lewis number characterize the diffusion in the simulations (see eqns.~\ref{eqn:rayleigh} \& \ref{eqn:lewis}).  The resolution is the number of Fourier or Chebyshev modes used in each direction.  The CFL safety factor is listed along with our choice of timestepper. The overshoot number ${\rm Ov}$ measures the ratio of inertial to buoyancy forces in the overshoot region (see eqn.~\ref{eqn:Ov}). The Reynolds number describes the degree of turbulence in the simulation (see eqn.~\ref{eqn:reynolds}). The three columns after the Reynolds number are the heights at which $\Dturb=\alpha\kappa$, where $\alpha=1$, $0.3$, or $0$.  For comparison, in simulation R8, the bottom of the convection zone is $z_0=1.180$ and the height of neutral buoyancy is $z_{\rm nb}=1.116$.  The last column is the overshoot length (normalized to the pressure scale height $H$), defined as the distance between the bottom of the convection zone and the location where $\Dturb=0$.} \label{tab:sims}
{\centering
  \begin{tabular}{ccccccccccc}
    \hline
    Name & ${\rm Ra}$ & ${\rm Le}$ & Resolution & Timestepper/CFL & ${\rm Ov}$ & ${\rm Re}$ & $\Dturb=\kappa$ & $\Dturb = 0.3\kappa$ & $\Dturb=0$ & $L_{\rm ov}$\\
    \hline
    \hline
    R7 & $10^7$ & $1$ & $256^3$ & RK222$^a$/1.0 & $4\times 10^{-4}$ & 150 & 1.123 & 1.097 & 1.066 & 0.111 \\
    R8 & $10^8$ & $1$ & $256^3$ & RK222/1.0 & $2\times 10^{-4}$ & 329 & 1.122 & 1.102 & 1.080 & 0.101 \\
    R9 & $10^9$ & $1$ & $512^3$ & SBDF2$^b$/0.4 & $1\times 10^{-4}$ & 751 & 1.122 & 1.107 & 1.091 & 0.090 \\
    R7L3& $10^7$ & $10^{1/2}$ & $256^3$ & RK222/1.0 & $4\times 10^{-4}$ & 150 & 1.133 & 1.102 & 1.061 & 0.116  \\
    R8L3 & $10^8$ & $10^{1/2}$ & $256^{3c}$ & RK222/1.0 & $2\times 10^{-4}$ & 329 & 1.133 & 1.109 & 1.083 & 0.098 \\
    R7L10& $10^7$ & $10$ & $256^{3c}$ & RK222/1.0 & $4\times 10^{-4}$ & 150 & 1.145 & 1.111 & 1.063 & 0.114 \\
    \hline
    \multicolumn{6}{l}{\textsuperscript{a}\footnotesize{Second order, two-stage Runge-Kutta method \citep{ascher97}}} \\
    \multicolumn{6}{l}{\textsuperscript{b}\footnotesize{Second order semi-backward differencing \citep{wang08}.}} \\
    \multicolumn{6}{l}{\textsuperscript{c}\footnotesize{The passive scalar field is evolved at $512^3$.}}
  \end{tabular}
}
\end{table*}

\subsection{Passive Tracer Field}

The goal of this work is to estimate turbulent diffusivities associated with convective overshoot.  To do this, we solve for the evolution of a passive tracer field $c$
\begin{align}
\partial_t c - D \nabla^2 c = - \vec{u}\dot\grad c.
\end{align}
The tracer $c$ heuristically represents the fuel concentration, and $D$ is a proxy for chemical diffusivity (and is required for numerical stability).  The tracer $c$ satisfies zero flux boundary conditions on the top and bottom of the domain, so its volume integral is conserved.  It is initialized with
\begin{align}
c = \frac{1}{2}\left[1 - \tanh\left(\frac{z-0.8H}{\Delta z_{\rm fl}}\right)\right],
\end{align}
which corresponds to $c=0$ in the convection zone and $c=1$ below the buoyancy peak in the stable region.

\section{Results}
\label{sec:results}

\begin{figure}
\includegraphics[width=\columnwidth]{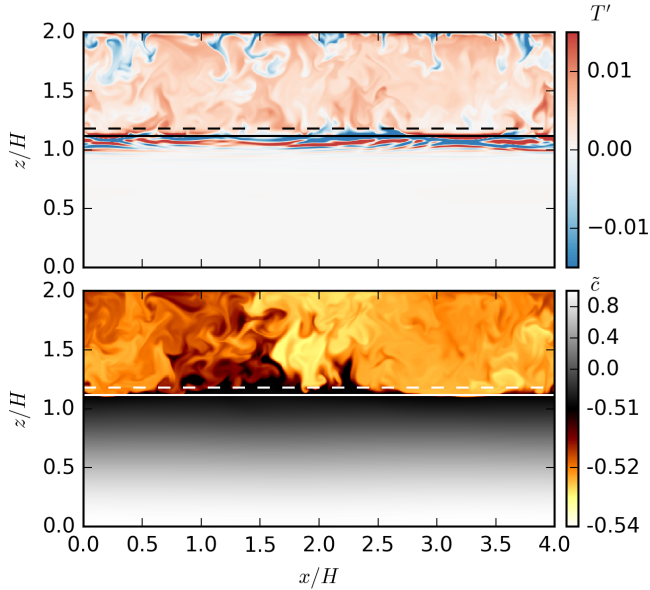}
  \caption{Two dimensional vertical slices of the temperature perturbation field (top) and the normalized passive scalar field (bottom) in simulation R9.  The color scale for $\tilde{c}$ consists of two linear maps, stitched together at $\tilde{c}\approx -0.5$ to show the small variations within the convection zone.  The dashed line shows the bottom of the convection zone, $z_0$, and the solid line shows $z_{\rm nb}$ the neutral buoyancy height. The perturbations below $z_{\rm nb}$ are waves and yield negligible mixing.}\label{fig:convection}
\end{figure}

After several convective turnover times, the system reaches a statistically steady state.  We visualize the convection in Fig.~\ref{fig:convection}, plotting 2D vertical slices of the temperature perturbation field and the normalized passive scalar field.  The temperature perturbation is $T' = T - \langle T \rangle_{x,y,t}$.  We normalize the passive scalar field by subtracting off the volume-average, and setting its value to 1 at the bottom boundary: 
\begin{equation}
\tilde{c} = \left(c - \langle c\rangle_{x,y,z}\right)/\left(\langle c(z=0)\rangle_{x,y}-\langle c\rangle_{x,y,z}\right)~.
\end{equation}

Fig.~\ref{fig:convection} includes dashed lines at the bottom of the convection zone, $z_0$, and solid lines at the height of neutral buoyancy $z_{\rm nb}$.  There is substantial convective overshoot between $z_0$ and $z_{\rm nb}$.  Below $z_{\rm nb}$, the buoyancy perturbations show the long, coherent structures of internal gravity waves.  These waves yield negligible mixing. 
  
\subsection{Self-Similar Solution}

\begin{figure}
\includegraphics[width=\columnwidth]{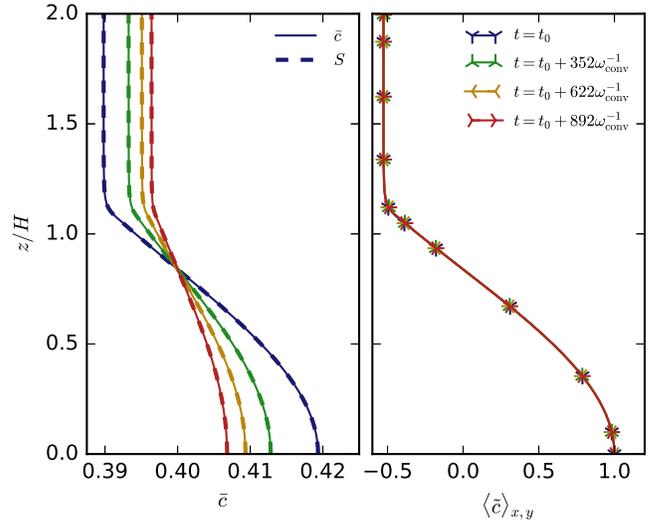}
  \caption{Horizontal average of the passive scalar field at four times in simulation R8.  $\bar{c}$ is also time-averaged around each time for $30\omega_{\rm conv}^{-1}$.  The passive scalar field is attracted to the self-similar solution, $C$ (right panel and equation~\ref{eqn:self similar}).  The left panel also shows the solution of the effective diffusion model (equation~\ref{eqn:diffusion model}).  The 1D effective diffusion model matches the 3D simulation.}\label{fig:c-bar}
\end{figure}

We now study the evolution of the horizontal average of the passive scalar field, $\bar{c}\equiv \langle c\rangle_{x,y}$.  After several convective turnover times, $\bar{c}$ approaches a self-similar solution.  The left panel of Fig.~\ref{fig:c-bar} shows the evolution of $\bar{c}$ in simulation R8, where $t_0$ is several turnover times after the beginning of the simulation.  The profiles collapse to a single curve after subtracting off the volume-average and normalizing the bottom value to unity (i.e., taking the horizontal average of $\tilde{c}$ shown in Fig.~\ref{fig:convection}).  This indicates that
\begin{align}\label{eqn:self similar}
\bar{c}(z,t)-\langle \bar{c}\rangle_{z}\rightarrow A(t)C(z),
\end{align}
where $A(t)$ is an amplitude, and $C(z)$ the vertical profile in the right panel of Fig.~\ref{fig:c-bar}.  Furthermore, we find that $A(t)=A_0\exp(-\lambda t)$.  $C$ thus satisfies the equation
\begin{align}
-\lambda C - D \partial_z^2 C = - \left\langle \vec{u}\dot\grad \frac{c}{A} \right\rangle_{x,y,t}
\end{align}
We now assume that the term on the right hand side can be written as a turbulent diffusion term.  This is the Fickian diffusion ansatz \citep[e.g.,][]{brandenburg09}.  The equation can be rewritten as
\begin{align}\label{eqn:effective diffusivity}
-\lambda C = \partial_z \left[ (D + \Dturb) \partial_z C\right],
\end{align}
where $\Dturb(z)$ is a turbulent diffusivity profile.  We can invert equation~\eqref{eqn:effective diffusivity} to solve for $\Dturb$ in terms of $\lambda$ and $C$ by integrating the equation with respect to $z$ and then dividing by $\partial_z C$.  We find that $\Dturb\ll D$ in the stable region, and is large $\sim w_{\rm rms}H_{\rm conv}$ in the convection zone; the value of $\Dturb$ is not well-constrained in the convection zone, as $\partial_zC$ is very close to zero. We find that the {\it effective} diffusivity, $D+\Dturb$ is well-fit by two error functions, one which varies from zero in the convection zone to $D$ in the stable region, the other which varies from zero in the stable region to $w_{\rm rms}H_{\rm conv}$ in the convection zone.  In the rest of this paper, we replace $\Dturb$ by a least-squares fit composed of these error functions.  Fig.~\ref{fig:diffusions} (left panel) includes both $|\Dturb|$ (dotted black line) and the least-squares fit (yellow line) for simulation R8.

\subsection{Turbulent Diffusivity Model}

\begin{figure}
\includegraphics[width=\columnwidth]{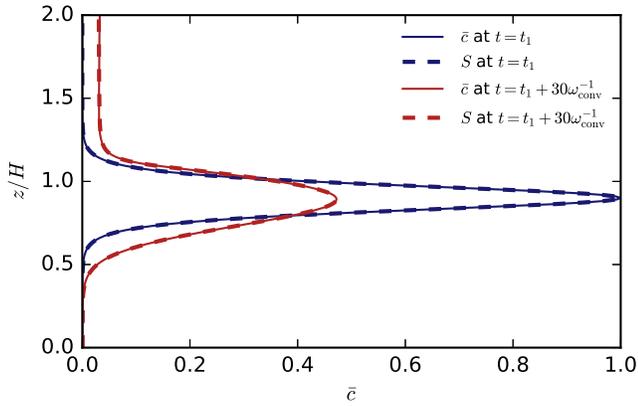}
  \caption{Horizontal average of a passive scalar field using the convection in simulation R8.  $c$ is initialized at $t_1$ to be horizontally uniform, with the vertical profile shown here.  The diffusion model equation~\eqref{eqn:diffusion model} was initialized with the same profile.  The 1D effective diffusion model matches the 3D simulation over the entire simulation.}\label{fig:diffusion model}
\end{figure}

To show that the convection acts like a turbulent diffusivity, we solve the model equation
\begin{align}\label{eqn:diffusion model}
\partial_t S(z,t) = \partial_z \left[(D+\Dturb)\partial_z S(z,t)\right].
\end{align}
If we initialize $S(z,t)$ with $\langle c(t=t_0)\rangle_{x,y}$ and use our fit for $\Dturb(z)$, we find that $S\approx A(t)C(z)$, as shown in Fig.~\ref{fig:c-bar}, for every simulation.

As a further test of the diffusion model, we re-initialized simulation R8 with a new concentration field profile halfway through the simulation at time $t_1$.  We solved equation~\eqref{eqn:diffusion model} with $S(z,t_1)=\bar{c}(t=t_1)$.  Fig.~\ref{fig:diffusion model} shows that $S\approx \bar{c}$ for the remainder of the simulation.

\subsection{Diffusion Profiles}\label{sec:diffusion profiles}

\begin{figure}
\includegraphics[width=\columnwidth]{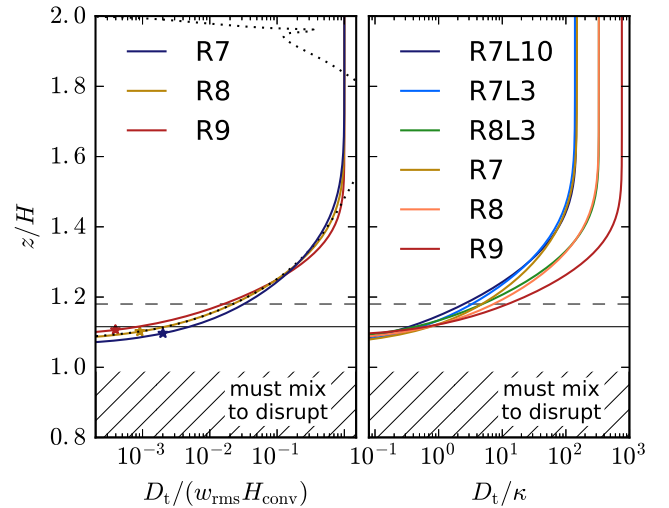}
  \caption{Turbulent diffusivity (equation~\ref{eqn:effective diffusivity}) as a function of height in each of our simulations, both in units of the characteristic convective diffusivity (left panel), and in units of the thermal diffusivity (right panel). We plot a fit to $\Dturb$ for all simulations, and also plot $|\Dturb|$ itself in the thin dotted line for simulation R8.  The dashed line shows the bottom of the convection zone, $z_0$, and the solid line shows $z_{\rm nb}$, the neutral buoyancy height.  In the left panel, the height at which $\Dturb=0.3\kappa$ is marked by an asterisk---mixing can only affect flame propagation above this point.  The hatched region shows the region that must be mixed in order to disrupt the flame (section~\ref{sec:flame-disruption}). Increasing ${\rm Ra}$ and/or ${\rm Le}$ causes $\Dturb$ to approach zero further away from the buoyancy peak, meaning that mixing is less significant for more realistic parameters.
  }\label{fig:diffusions}
\end{figure}

We plot the turbulent diffusion profiles $\Dturb(z)$ for each of our simulations in Fig.~\ref{fig:diffusions}, both in units of the characteristic convective diffusivity (left panel), and in units of the thermal diffusivity (right panel). In the convection zone, the diffusivity is about equal to the convective diffusivity, partially dictated by our choice of fit.  The turbulent diffusivity drops from its convective value {\it within} the convection zone.  This cannot be attributed to the change in the horizontal average of $w^2$ near $z_0$ \citep[similar to][]{Jones16}.  Deep within the stably stratified region, the turbulent diffusivity is nearly zero.

We are interested in how $\Dturb$ transitions from large values in the convection zone to small values in the stable region.  In this respect, the behavior of $\Dturb/\kappa$ is very similar in all simulations (Fig.~\ref{fig:diffusions}, right panel).  Heuristically, we expect mixing to play a role in the propagation of flames when $\Dturb\sim\kappa$.  We find that the height at which $\Dturb=\kappa$ is almost independent of ${\rm Ra}$, but increases with ${\rm Le}$ (Table~\ref{tab:sims}), i.e., moves closer to the convective boundary and further from the ``flame.''  In section~\ref{sec:flame-disruption}, we find that a more precise criterion for flame disruption is $\Dturb\gtrsim0.3 \kappa$ in the region in which $N \gtrsim 0.1 N_{\rm fl}$.  The height at which $\Dturb=0.3\kappa$ increases with both ${\rm Ra}$ (Fig.~\ref{fig:diffusions}, left panel), and ${\rm Le}$ (Table~\ref{tab:sims}), suggesting that flame disruption becomes less likely for more realistic values of ${\rm Ra}$ and ${\rm Le}$

Two common parameterizations of convective overshoot are exponential overshoot, in which the turbulent diffusivity drops exponentially with distance from the end of the convection zone \citep[e.g.,][]{Herwig00}, and an overshoot length, in which the convective diffusivity is set to zero at a length $L_{\rm ov}$ beyond the convection zone \citep[e.g.,][]{Shaviv73, Maeder75}.  In all our simulations, $\Dturb$ is {\it negative} below a critical height (although the {\it effective} diffusivity $D+\Dturb$ is everywhere positive).  This suggests that a good parameterization of our simulations would be an overshoot length, rather than exponential overshoot.  We define the overshoot length $L_{\rm ov}$ to be the distance between the bottom of the convection zone (where $N^2=0$), and the location where $\Dturb=0$, and report it in Table~\ref{tab:sims}.  All lengths in the paper, including $L_{\rm ov}$ are normalized to the pressure scaleheight $H$.    Below the point at which $\Dturb=0$ the absolute value of $\Dturb$ is very small.\footnote{We cannot place strong constraints on $|\Dturb|$  when its value is very small, as its value can be influenced by some combination of: 1.~Timestepping errors due to using a low ($2^{\rm nd}$) order timestepper; or 2.~Errors in the calculation of $C(z)$ or $\lambda$ due to insufficient averaging.}

The weak dependence of $\Dturb$ in the overshoot region on the diffusivities of the system suggests that the height at which $\Dturb=0.3\kappa$ and the overshoot length $L_{\rm ov}$ are determined primarily by the length scale on which the buoyancy frequency profile changes from zero to order $\omega_c$, rather than a diffusive length scale.  This suggestions that the key lengthscale in the problem is $\sim z_{\rm nb}-z_0$ (see Fig.~\ref{fig:buoyancy frequency}).  Indeed, the overshoot lenght $L_{\rm ov}\sim z_{nb}-z_0$ in all of our simulations (Table~\ref{tab:sims}).  This is because dense plumes falling through the convection zone become much lighter than their surroundings below $z_{\rm nb}$, so they cannot penetrate much further to produce mixing within the flame.  We expect the overshoot length to scale as
\begin{align}
L_{\rm ov} - (z_0-z_{\rm nb}) \sim {\rm Ov}^{1/3}.
\end{align}
Our simulations do not explore a sufficiently wide range of ${\rm Ov}$ to test this scaling.  Although increasing ${\rm Ra}$ or ${\rm Le}$ further will introduce smaller eddies into simulations, we do not believe these smaller eddies will enhance mixing, as they are subject to the same buoyancy barrier as the larger plumes resolved in the simulations presented here.

\subsection{Flame Disruption in MESA}
\label{sec:flame-disruption}

We explore the secular effects of mixing on flame propagation via a
series of numerical experiments using MESA.  We begin with the
evolution of a $\unit[9.5]{\Msun}$ star (the same calculation
discussed in Section~\ref{sec:carb-flame-prop}).  We save a model when
the carbon flame is at a Lagrangian mass coordinate of
$\unit[0.2]{\Msun}$.  We load this model in revision 8118 of MESA and
use the built-in \texttt{other\_D\_mix} routine to introduce an
artificial chemical diffusivity in the vicinity of the flame.  We then
observe whether this additional mixing affects the behavior of the
flame.  In the absence of additional mixing, the carbon-burning
luminosity in the flame is smooth (in time) and roughly constant, with
some secular variation as the flame propagates inward.  We evolve the
MESA models for $\approx \unit[2000]{yr}$,
which is $\approx 10$ self-crossing times for the flame; in this time, the
unperturbed flame propagates inwards through
$\approx \unit[0.1]{\Msun}$ of material. We classify the flame as
``disrupted'' if the carbon-burning luminosity decreases significantly
(by more than a factor of $\approx$ 10) or exhibits oscillatory
behavior (by more than $\approx$ 10 \%).

First, we set the chemical diffusivity ($D_t$) roughly
equal to the convective diffusivity, $\unit[10^{12}]{cm^2\,s^{-1}}$
(which is $\sim H^2\omega$), in the region of the flame where
$N<N_{\rm crit}$.  This allows us to determine the region where
significant mixing is required to disrupt the flame.  Increasing
$N_{\rm crit}$ increases the amount of material in which additional
mixing occurs, similar to increasing the overshoot length
scale.\footnote{However, unlike overshooting, the mixing that we
  introduce is not spatially tied to the convective boundary.}  We
find the flame is only disrupted if
$N_{\rm crit} \gtrsim 0.3 N_{\rm fl}$, where $N_{\rm fl}$ is the peak
of the buoyancy frequency.  This reflects the fact that
it is necessary to mix material in the
region where the bulk of the nuclear energy release is occurring in
order to disrupt the flame.

Second, we set the chemical diffusivity to be a constant factor times
the thermal diffusivity over a region where $N<N_{\rm crit}$. This
allows us to determine the ratio $\Dturb/\kappa$ needed to disrupt a
flame.  In terms of the opacity $\kappa_\star$, the thermal
diffusivity is given by
\begin{equation}
  \kappa = \frac{4 a c T^3}{\kappa_{\star} \rho^2 c_{\mathrm{P}}}
\end{equation}
where $a$ is the radiation constant, $c$ the speed of light, $T$ the
temperature, $\rho$ the density, and $c_{\mathrm{P}}$ the specific
heat at constant pressure.  For a value of
$N_{\rm crit} = 0.3 N_{\rm fl}$, we find that the flame is only
disrupted if $D_t > 0.3 \kappa$.  This agrees with our heuristic
that $\Dturb\sim \kappa$ is necessary for flame disruption.
If the mixing is allowed to be even deeper into the flame (higher $N_{\rm crit}$), lower diffusivities are required; however, because our simulations suggest the turbulent diffusivity drops off very sharply with depth, we believe the most germane requirement for flame disruption is that from the shallowest mixing.

We use the criteria derived from these MESA calculations to interpret
the results of our Dedalus simulations.  The Dedalus
simulations address where and how efficiently convection mixes
material in the presence of a buoyancy barrier.  However, because they
do not self-consistently model a conductively-propagating flame, they
cannot directly answer the question of whether a flame disrupts.  The MESA
calculations directly address whether convective mixing with a
specific efficiency (relative to $\kappa$) and at a specific location
(relative to $N$) is sufficient to disrupt a flame.  We show these
criteria in Fig.~\ref{fig:diffusions}: the region where
$N > 0.3 N_{\rm fl}$ is hatched and the points where
$\Dturb = 0.3\kappa$ are marked with stars.  In all our Dedalus
simulations, the stars are outside the hatched region,
which implies that the mixing
observed in Dedalus would not be sufficient to disrupt the flame.

\section{Conclusions}
\label{sec:conclusions}

This paper describes simulations of an idealized model of convectively bounded carbon flames.  The simulations are in the Boussinesq approximation, and assume a Brunt-V\"{a}is\"{a}l\"{a} frequency profile motivated by MESA simulations of carbon flames (Fig.~\ref{fig:buoyancy frequency}).  On the convective timescale, carbon flames are almost stationary, so we do not explicitly include any nuclear burning in our model.

The simulations evolve a passive scalar field which heuristically represents the carbon species fraction.  Overshooting plumes mix the passive scalar into the convection zone.  The passive scalar field quickly approaches a self-similar solution (equation~\ref{eqn:self similar}; see Fig.~\ref{fig:c-bar}), allowing us to calculate an effective diffusivity profile $\Dturb(z)$.  The horizontally averaged 3D evolution of the passive scalar field is very well approximated by the solution of a 1D diffusion equation (equation~\ref{eqn:diffusion model}; see Fig.~\ref{fig:diffusion model}).

Our simulations have large diffusivities compared to real stars.   Despite the unphysical parameter regime of our simulations, we believe that we can still draw strong conclusions about mixing in real carbon flames, because of the clear trends in the simulation results as the parameters become more realistic, i.e., with increasing Rayleigh and Lewis numbers.

Carbon flames have $\kappa/D \sim 10^6$, but convective mixing can stall a flame if the turbulent mixing due to overshoot is such that $\Dturb\sim\kappa$ within the flame.   Overshoot in 1D stellar models is sometimes modeled by exponentially decreasing the diffusion coefficient outside the convection zone over a characteristic length \citep[e.g.,][]{Herwig00}.  This parameterization does not in fact apply to our simulations, which have turbulent diffusivities which decrease as Gaussians, and then become negative below a critical height (Sec.~\ref{sec:diffusion profiles}).  This suggests that a more useful parameterization is an overshoot length, as we find no convective mixing below a critical height.

MESA calculations suggest that a region near the peak of the buoyancy frequency ($N\sim 0.3N_{\rm fl}$) must be mixed with $\Dturb>0.3 \kappa$ in order to disrupt the flame (Sec.~\ref{sec:flame-disruption}). None of our simulations of convective overshoot show any convective mixing in this region.  In all of our simulations, the height at which $\Dturb=0.3 \kappa$ is well outside the region near the peak of the buoyancy frequency that MESA simulations show must be mixed in order to stall the flame (Fig.~\ref{fig:diffusions}).  Moreover, this height shifts closer and closer to the convection zone (away from the flame) as either the Rayleigh number or $\kappa/D$ (the Lewis number) increase towards more realistic values.

Furthermore, our simulations greatly overestimate the mixing efficiency, as our buoyancy frequency increases only modestly with depth (Fig.~\ref{fig:buoyancy frequency}).  Although the ratio of inertia in our convective plumes to the stabilizing buoyancy force is very small ($\sim 10^{-4}$; see Table~\ref{tab:sims}), we estimate that our simulated plumes are nonetheless more powerful than realistic plumes by a factor of at least $\sim 10^6$.

Taken together, these results strongly suggest that convection provides insufficient mixing to disrupt real carbon flames.  The only way out of this conclusion is to posit that for yet higher ${\rm Ra}$ or ${\rm Le}$ numbers, the trends we find in mixing with increasingly realistic parameters reverse.  Although we cannot rule this out, we regard it as unlikely.  Physically, the lack of mixing is due to a simple physical principle: convective plumes must overcome a huge buoyancy barrier to reach the flame.    There is no reason to expect them to suddenly be able to do so at even higher ${\rm Ra}$ or ${\rm Le}$.   As a result, we conclude that convection provides insufficient mixing to disrupt a carbon flame and that ``hybrid C/O/Ne'' WDs are unlikely to be a typical product of stellar evolution.

We have neglected important physics in this work, including rotation, magnetism, density stratification, and nuclear burning.  However, it seems difficult for these effects to overcome the potential energy barrier, so we do not believe they will change our conclusion.

Internal gravity waves generated by the convection could mix the fluid via breaking.  The wave amplitude increases as $\sqrt{N}$ as the waves leave the convection zone and approach the flame.  Waves can break if $k_r\xi_r\sim 1$, where $\xi_r$ is the vertical displacement and $k_r$ is the vertical wavenumber.  Neglecting damping, theoretical models of internal wave generation by convection \citep[e.g.,][]{lecoanet2013} claim $k_r\xi_r\sim 1$ at the peak of the buoyancy frequency, $N_{\rm fl}$.  However, the waves linearly damp due to thermal diffusion (which does not lead to chemical mixing).  For carbon flames, we estimate the linear damping to become important near $N_{\rm fl}$, so it is unclear if the waves would break. Furthermore, breaking waves may only mix the unburnt fuel near $N_{\rm fl}$, having little effect on flame propagation.

Our simulations all have $\nu=\kappa$, but in stars, we estimate the Prandtl number ${\rm Pr}=\nu/\kappa\sim 10^{-5}$.  Thus, there are small-scale motions which are isothermal, but not strongly influenced by viscosity.  These motions can penetrate the buoyancy gradient in the flame, and thus are expected to enhance mixing.  At a fixed ${\rm Pr}$, we expect mixing to become less efficient as ${\rm Ra}$ increases, as the length scale on which perturbations are isothermal will decrease.  Thus, as ${\rm Ra}$ increases, there will be less and less energy in isothermal perturbations.

More quantitatively, the largest length scale for isothermal perturbations is $\ell\sim\kappa/v_{\ell}$, where $v_{\ell}$ is typical velocity of eddies of size $\ell$.  Assuming a Kolmogorov cascade with $v_{\ell}\sim \omega_0 H (\ell/H)^{1/3}$, we have $v_{\ell}\sim \unit[3\times10^{2}]{cm\,s^{-1}}$ and $\ell\sim\unit[10]{cm}$.  The diffusive mixing produced by these eddies is about $\Dturb\sim\ell v_{\ell}\sim \kappa$, which is enough to disrupt the flame.  However, these eddies will travel a depth $\ell\ll \delta$, and thus should not penetrate far enough into the flame to disrupt it.  Future work should validate these estimates.

Given the strong intermittency of convective turbulence, it is also possible that the majority of overshoot mixing may be caused by a few rare but powerful plumes.  Although our study cannot rule out this possibility, we note that there are about $\sim 10^6$ convective turnover times in the lifetime of a carbon flame.  This is many fewer turnover times than in other astrophysical contexts (e.g., the solar convection zone), so rare events may be less important for carbon flames.

Future work should also study mixing via overshoot in oxygen-neon flames,
which is important for understanding whether stars at the top of the
SAGB mass range undergo Fe core collapse or electron-capture-induced
ONe core collapse \citep{Jones14}.

\section*{Acknowledgments}

\noindent{}We thank two anonymous referees for elucidating comments.  We acknowledge stimulating
workshops at Sky House where these ideas germinated.  D.L.~is supported by the Hertz Foundation.  E.Q.~is supported in part by a Simons Investigator Award from the Simons Foundation. J.S.~is supported by NSF grant AST 12-05732.  G.M.V.~acknowledges support from the Australian Research Council, project number DE140101960. J.S.O.~is supported by a Provost's Research Fellowship from Farmingdale State College.  This research is funded in part by the Gordon and Betty Moore Foundation through Grant GBMF5076 to L.B.~and E.Q.  Resources supporting this work were provided by the NASA High-End Computing (HEC) Program through the NASA Advanced Supercomputing (NAS) Division at Ames Research Center.  This project was supported by NASA under TCAN grant number NNX14AB53G.   This work was partially supported by the National Science Foundation under grants PHY 11-25915 and AST
12-05574.

\end{document}